\documentclass{ws-ijmpcs}

\begin{document}

\markboth{Elena Amato}
{The theory of Pulsar Wind Nebulae}

%%%%%%%%%%%%%%%%%%%%% Publisher's Area please ignore %%%%%%%%%%%%%%%
%
\catchline{}{}{}{}{}
%
%%%%%%%%%%%%%%%%%%%%%%%%%%%%%%%%%%%%%%%%%%%%%%%%%%%%%%%%%%%%%%%%%%%%

\title{THE THEORY OF PULSAR WIND NEBULAE}

\author{ELENA AMATO}

\address{INAF - Osservatorio Astrofisico di Arcetri, Largo E. Fermi, 5, I-50125, Firenze, Italy\\
amato@arcetri.astro.it}

\maketitle

\begin{history}
\received{10 November 2013}
\revised{10 November 2013}
\end{history}

\begin{abstract}
I will review the current status of our theoretical understanding of
Pulsar Winds and associated nebulae (PWNe). In recent years,
axisymmetric models of pulsar winds with a latitude dependent energy
flux have proved very successful at explaining the morphology of PWNe as
seen in the X-rays. This success has prompted developments aimed at
using multi-wavelength observations of these nebulae as a diagnostics of
the hidden physics of the pulsar wind and of the mechanism(s) through
which particles are accelerated in these sources.

I will discuss these most recent developments in terms of the
information that we infer from detailed comparison of
simulated non-thermal emission with current
observations. 

\keywords{ISM: supernova remnants; MHD; acceleration of particles; }
\end{abstract}

\ccode{PACS numbers: 98.38.Mz,98.38.Fs,96.50.Pw}

\section{Introduction}	
Pulsar Wind Nebulae (PWNe) are a special class of Supernova Remnants (SNRs). Historically they have been defined based on their observational properties, namely a center filled emission morphology, a flat spectrum at radio wavelengths, and an overall very broad spectrum of non-thermal emission, extending from the radio band to high energy $\gamma$-rays. 
The ultimate source of energy behind the emission is a fast-spinning neutron star, usually also observed as a pulsar. Most of the rotational energy lost by such a star goes into the launching of a highly relativistic magnetized wind, primarily made of electron-positron pairs. In the case of young pulsars, the wind is surrounded by the debris of the supernova explosion, which are in expansion at a much lower (non-relativistic) velocity. When the wind first impacts on the SNR, a reverse shock is launched towards the pulsar. At this shock (the termination shock, TS hereafter) the wind is slowed down and its bulk energy transformed into that of the magnetized, relativistically hot plasma responsible for the nebular emission. 

Before discussing in more detail the current status of our understanding of such systems, it seems appropriate to put them in the general contest of high energy astrophysics and briefly review the reasons why these sources are of interest for several different fields of physics and astrophysics. 

The first obvious reason of interest for PWNe comes from the fact that they provide possibly the best window on pulsar physics. While pulsars are thought to be the primary leptonic antimatter factories in the Galaxy, the exact amount of pair-production in their magnetospheres, the so-called pair multiplicity, is not well established. PWNe shine with luminosities that are tens of percent of the pulsar total spin-down power $\dot E$, a much larger fraction than what comes out in the form of pulsed emission in any frequency band.
Modeling of PWNe provides the tightest constraints on the subject of pulsar multiplicities \cite{arons12,bucciantini11}, in a time when assessing the role of pulsars as positron producers seems more urgent than ever, after the detection of the so-called "positron excess" by PAMELA \cite{pamela} and AMS02 \cite{ams02}. The latter might be associated with dark-matter related processes, but only after subtraction of all relevant astrophysical backgrounds, of which pulsars are likely the main contributors \cite{serpico12}.

Understanding PWNe is important from the point of view of cosmic ray physics also in other respects: in these sources we observe the workings of the most relativistic shocks in Nature, an extreme version of the shocks that are usually invoked as particle accelerators to the highest energies. In PWNe we have direct evidence of particle acceleration up to PeV energies, around the maximum we think achievable by galactic accelerators and the acceleration efficiency is extremely high, tens of percent of the pulsar $\dot E$.
All this is achieved in an environment, that of a magnetized relativistic shock, where particle acceleration is most difficult to understand. Being so close and bright, PWNe are the place to start from in the quest for understanding particle acceleration at relativistic shocks and many other aspects of the physics of relativistic sources.

\section{The MHD description of PWNe}
\label{sec:mhd}
Most of what we know about PWNe actually comes from the study of just one object, the Crab Nebula, the class prototype. The Crab Nebula was the first source of non-thermal radiation ever identified in the sky and is one of the best studied object in the Universe (second only to the Sun): since discovery, it has been observed by virtually every instrument ever built and the wealth of data is absolutely extraordinary in comparison with any other relativistic source. From the Crab Nebula we learnt that PWNe owe their luminosity to synchrotron (and, at high frequencies, also Inverse Compton) emission by relativistic particles gyrating in a magnetic field about 100 times more intense than that in the ISM. The source of the nebular luminosity was suggested to be a fast-rotating, highly magnetized neutron star \cite{pacini67} even before the discovery of pulsars, and the identification of one of these stars in the heart of the Crab Nebula.

In spite of the fact that pulsar physics is still rich of open questions, especially concerning the emission of pulses and the detailed mechanisms of pair production, nowadays we have a reasonably well established description of the structure of the wind that emanates from such objects. The wind must be mostly made of pairs and is magnetically dominated at the origin. If we define the magnetization parameter, $\sigma$, as the ratio between Poynting flux and ram pressure of the wind ($\sigma=B^2/(4\pi n m_e \Gamma^2 c^2)$, with $n$ the comoving particle number density, $m_e$ the electron mass and $\Gamma$ the wind Lorentz factor), the previous statement implies $\sigma \gg 1$ at the base of the wind. We also know that the wind is "dense", namely that the flux of pairs is much larger than what can be directly extracted from the star, the so-called Goldreich \& Julian's pair injection rate \cite{GJ69}, $\dot N_{GJ}\approx 3 \times 10^{30} \mu_{30}P^{-2}\ s^{-1}$, with $P$ the pulsar period and $\mu$ its magnetic moment in units of $10^{30} G\ {\rm cm}^3$. This latter fact is deduced from observations of synchrotron and ICS radiation, that point to the presence of a large number of pairs in PWNe: at least at the highest energies these pairs can only come from the wind. High $\sigma$ and dense wind, support the idea that a Force-Free (FF) description of the wind is viable. 

In the last few years, a few different groups have performed numerical simulations of pulsar winds within the framework of both FF and relativistic MHD, and more recently even dissipative FF \cite{spitkovsky06,kalapo12,tchekho13}, finding similar results. The emerging structure of the wind is in excellent agreement with the predictions of the analytic split monopole model \cite{michel73}: the streamlines of the outflow become asymptotically radial beyond the light cylinder radius ($R_{LC}=2 \pi c P$) and the magnetic field becomes predominantly toroidal. The energy flux has a latitude dependence which is that of the Poynting flux at large distances, $F\propto \sin^2 \theta$. At low latitudes around the equator, within an angular sector of width equal to twice the inclination between the magnetic and rotation axis of the pulsar, an oscillating current sheet develops, which provides a very plausible location for magnetic dissipation. 

Even before numerical simulations confirmed the split monopole solution as a good representation of the wind structure, the $\sin^2 \theta$ dependence of the energy flux from the pulsar had already been implied as a possible explanation of the X-ray morphology of PWNe. Early after the launch of the {\it Chandra} X-ray telescope in 2000, the Crab Nebula, and a number of other PWNe were found to show a very intriguing jet-torus morphology. The observation of jets was particularly puzzling, because the fact that their origin was so close to the pulsar seemed to require collimation upstream of the TS. In an environment such as that of PWNe the most obvious candidate mechanism for collimation is magnetic hoop stress, but this is very inefficient in a ultrarelativistic flow \cite{lyub02} . On the other hand, an anisotropic energy flux from the pulsar, as early suggested \cite{lyub02,bogo02}, causes the shock to be highly oblate, closer to the pulsar along the rotation axis than at the equator. 2D axisymmetric relativistic MHD simulations, performed by several different groups \cite{KL04,ldz04,bogo05} have proved that if the wind magnetization is sufficiently high, collimation is then ensured post-shock: the magnetic hoop stresses build up in the downstream, and for high enough $\sigma$ ($\sigma>0.01$ in the case of the Crab Nebula) the field reaches equipartition in the Nebula and at that point its tension diverts the flow towards the polar axis, giving rise to a jet. 

\section{Open questions}
\label{sec:questions}
The energy outflow from the pulsar can be written as:
\begin{equation}
\dot E_R=\kappa \dot N_{GJ} m_e c^2 \Gamma \left(1+\frac{\xi m_i}{\kappa m_e}\right) \left(1+\sigma \right)
\label{eq:pwout}
\end{equation}
where $m_i$ is the mass of ions, $\xi$ is a number between 0 and 1 defining their extraction rate in terms of $\dot N_{GJ}$ and $\kappa$ is the pulsar multiplicity. We estimate $\dot E_R$ and $\dot N_{GJ}$ knowing the pulsar period. All other parameters appearing in Eq.~\ref{eq:pwout} can only be estimated from modeling the PWN emission.

For many years the main puzzle in PWN physics has resided in the so-called $\sigma$ problem. The problem is the following. We just mentioned that $\sigma \gg1$ at the base of the wind. Within ideal MHD $\sigma$ cannot change, but 1D modeling of PWNe within ideal relativistic MHD \cite{kc1} was found to constrain $\sigma<<1$ at the wind TS: $\sigma \approx v_N/c$ with $v_N$ the expansion velocity at the outer boundary of the nebula (that of the SN ejecta). Axisymetric MHD simulations require a larger value of $\sigma$ in order for jets to appear \cite{ldz04}, but yet a value that is much less than unity, hence still leaving the problem of achieving efficient conversion of the wind magnetic energy into particle kinetic energy basically unaltered.

Another important open question concerns the presence and possible physical relevance of ions in the wind. In principle one could expect ions (protons or Fe nuclei) to be extracted from the star surface at a rate $\dot N_{GJ}$ and then become part of the wind. If the wind Lorentz factor is of order $few \times 10^6$, and the pulsar multiplicity $\kappa \approx 10^4$, as inferred from early 1D MHD modeling \cite{kc2} and required for reproducing the optical/X-ray spectrum of the Crab Nebula, then these would be PeV ions and dominate the energy content of the wind, in spite of being a minority by number. Clearly the situation is completely different if $\Gamma \approx 10^5$ and $\kappa \approx 10^6$ as required to explain the radio emission as due to particles continuously injected in the nebula \cite{bucciantini11}. In this case, even ions injected at a rate $\dot N_{GJ}$ would be energetically irrelevant.

All these parameters, $\xi$, $\kappa$ and $\sigma$, apart from holding important information on the inner workings of pulsar magnetospheres, are also crucial, as will be later discussed in more detail, for nailing down the acceleration mechanism(s) at work in these sources. In the following I discuss some recent results in terms of constraining the values of $\kappa$ and $\sigma$ with the help of relativistic MHD simulations, and implications for particle acceleration. 

\section{Investigating the wind magnetization}
\label{sec:sigma}
The best way to learn about the wind is detailed modeling of PWN dynamics and emission properties. 
In order to learn about the wind, detailed modeling of PWN dynamics and emission properties is required. This has been done for the past few years within the framework of 2D axisymmetric MHD models supplemented with tracers for the evolution of accelerated particles, that are subject to adiabatic and synchrotron losses while propagating away from the acceleration side, presumably the TS \cite{ldz06,volpi08}.

Among the unknown parameters of the pulsar wind, the magnetization is the only one that can be constrained based on the dynamics and morphology alone: for given ejecta and a given value of the pulsar spin-down luminosity, $\sigma$ governs the size of the TS. 
\begin{figure}
\includegraphics[scale=.2]{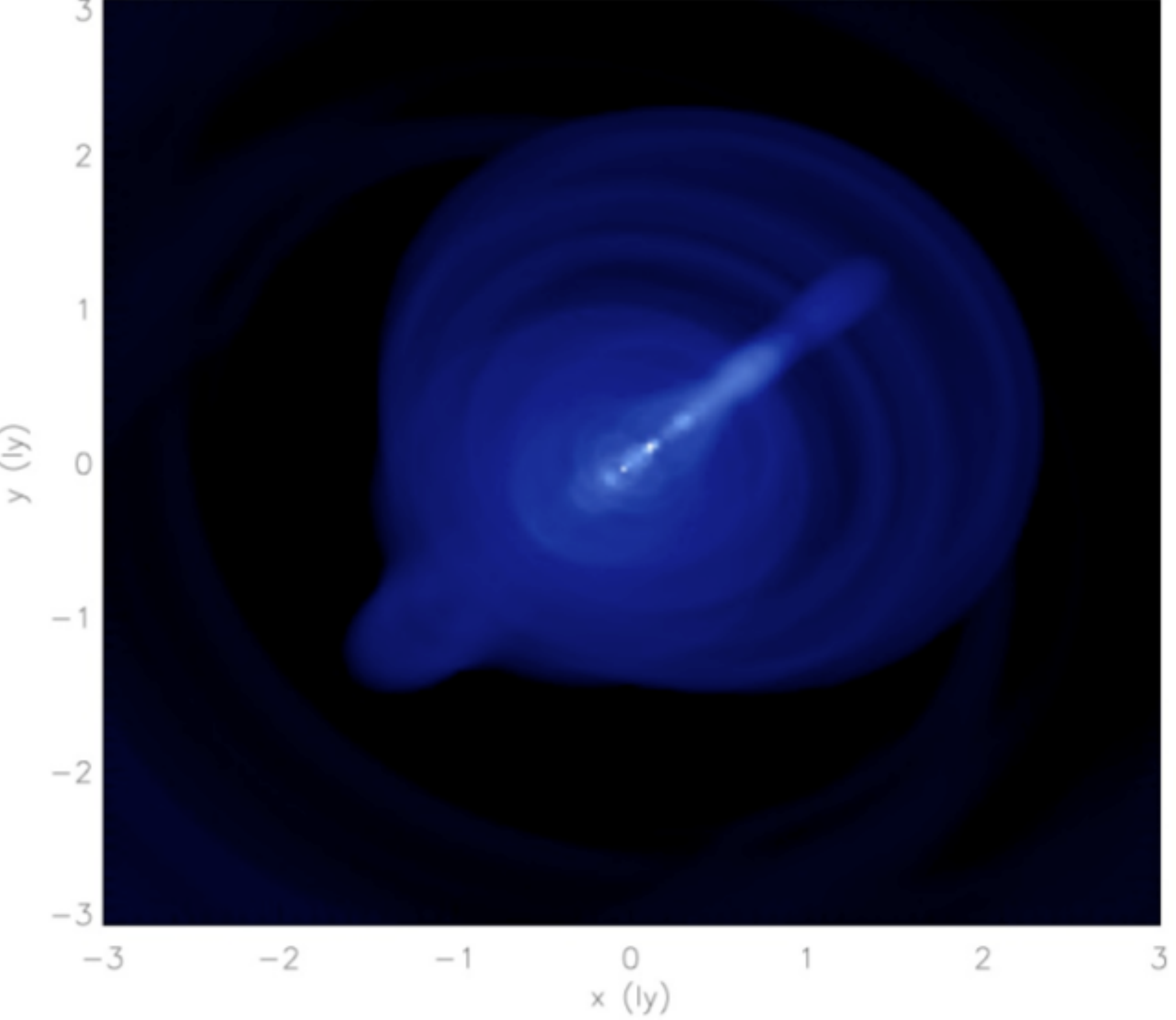}
\includegraphics[scale=.2]{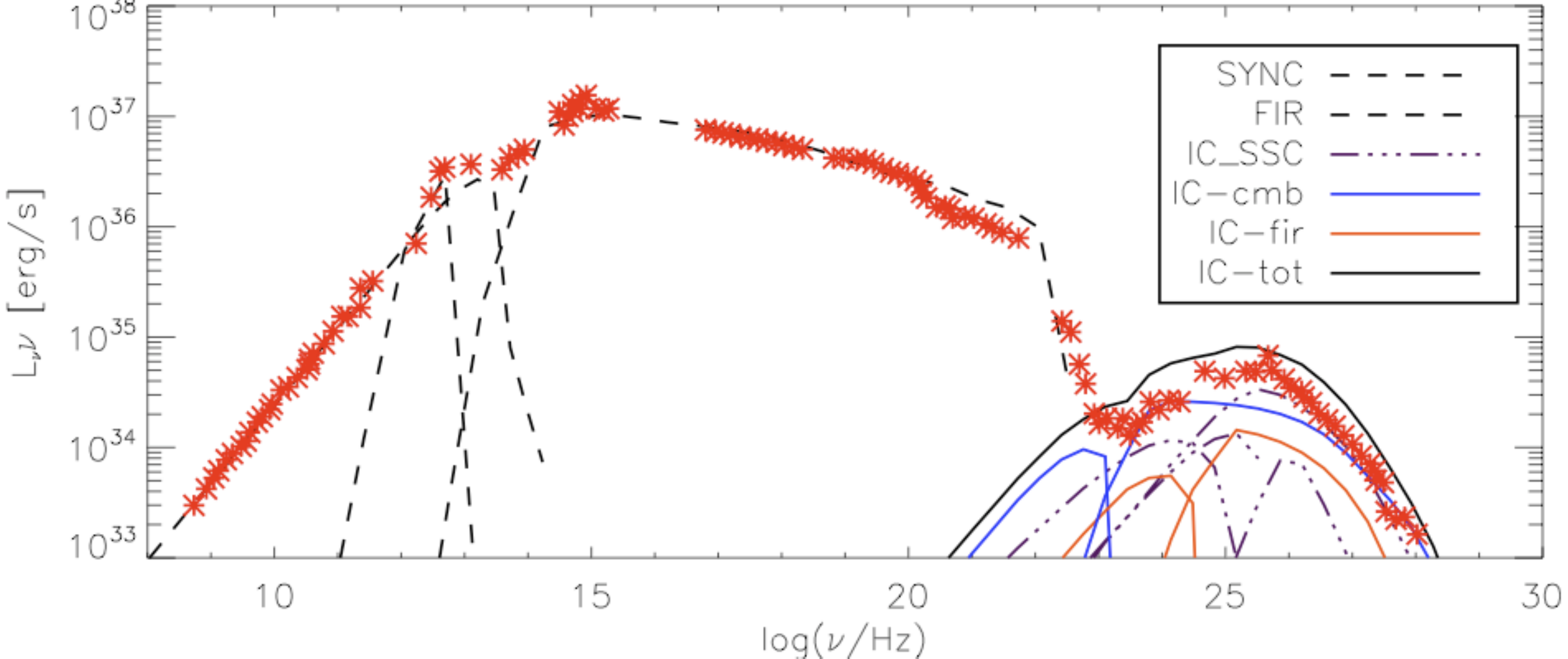}
\caption{Left panel: X-ray emission map of the Crab Nebula computed for $\sigma=1.5$. Right panel: corresponding integrated emission spectrum.}
\label{fX}
\end{figure}
The models that best reproduce the high energy emission from the Crab Nebula correspond to $\sigma \approx 0.03$, which is 10 times larger than the value inferred from 1D modeling, but still much below unity. 
These models, however, while reproducing the morphology down to very fine details, have problems accounting for the spectral properties of the Crab Nebula. The resulting average magnetic field in the simulated nebula at its current age is $B_{sim}\approx 50 \mu G$, roughly a factor 3-4 lower than estimated in reality \cite{ataha96}. As a consequence, the synchrotron emission can only be reproduced by assuming a correspondingly larger number of emitting particles, and with an artificially soft spectrum ($N(E) \propto E^{-\alpha_p}$ with $\alpha_p=2.8$ rather than the 2.2 traditionally inferred \cite{volpi08}), in order to compensate for the lack of synchrotron steepening due to the low $B_{sim}$. But this artificial enhancement of the number of particles then clearly shows up when one computes the part of the spectrum that derives from Inverse Compton scattering: the gamma-ray emission is largely over-predicted. 

In terms of integrated emission this discrepancy is much reduced if a higher $\sigma$ is adopted: for $\sigma\approx 2$ the integrated emission at all frequencies can be reasonably reproduced. But the price to pay is that the morphology is now very different: the size and brightness of all the low latitude features is very much reduced and the nebula appears as just a jet (see Fig.~\ref{fX}).

One possible solution to reconcile morphology and spectrum has been recently explored by Ref.~\refcite{porth13}, that showed that in 2D and 3D the same value of $\sigma$ does not correspond to the same morphology. The proposal was initially made in Ref.~\refcite{begelman98} that a possible solution to the $\sigma$ paradox could come from the development of kink instabilities (only allowed in 3D) that would tangle the magnetic field so as to reduce the associated hoop stress. Indeed the recent simulations \cite{porth13} fully confirm this idea, except for the fact that not only the development of an important poloidal component of the magnetic field is observed, but efficient magnetic dissipation also takes place. This fact leaves the question open of whether all problems at accounting for both the morphological and spectral properties of the Crab Nebula are really solved, once the results of the simulation which only lasts for less than 1/10 of the age of the nebula, are extrapolated to the current time.   

In any case, even just considering the emission maps that can be built on top of axisymmetric simulations, one concludes that the morphology is best explained assuming that the magnetic field becomes progressively tangled: with a purely azimuthal magnetic field the torus is too faint compared with the inner ring and all axisymmetric structures show emission profiles that decline too fast moving away from the line of sight along the azimuthal direction \cite{volpi08}.

\section{Investigating the pulsar multiplicity}
\label{sec:kappa}
I mentioned that ambiguity in the estimate of the pulsar multiplicity based on observations and modeling of PWNe are primarily associated with the fact that we do not know whether the radio emitting particles, the dominant ones by number, are part of the current pulsar outflow. In the case of the Crab Nebula the estimated $\kappa$ changes by at least two orders of magnitude depending on whether the low energy emitters are part of the wind. This has also consequences on the composition of the wind: if $\kappa$ is as high as required from radio emission then ions, even if present, cannot be energetically dominant, a fact that in turn makes a selection among the suggested acceleration mechanisms at the shock (see below).

In order to investigate the origin of the radio particles, Ref.~\refcite{olmi13} have considered different scenarios for their acceleration and consequent spatial distribution, and checked whether it is possible to discriminate among them based on the emission morphology. While a scenario in which low energy particles are pure fossile, namely injected only for a short time after the SN explosion, with no further reacceleration, can actually be excluded \cite{olmi13}, it does not seem possible to discriminate between two other scenarios, namely ongoing injection at the TS followed by advection with the flow (case A), and uniform distribution in the nebula (case B). The latter description should correspond to the physical picture in which low energy particles are continuously accelerated (or reaccelerated) by spatially distributed turbulence in the body of the nebula (Fermi II type mechanism).

\begin{figure}
\includegraphics[scale=.38]{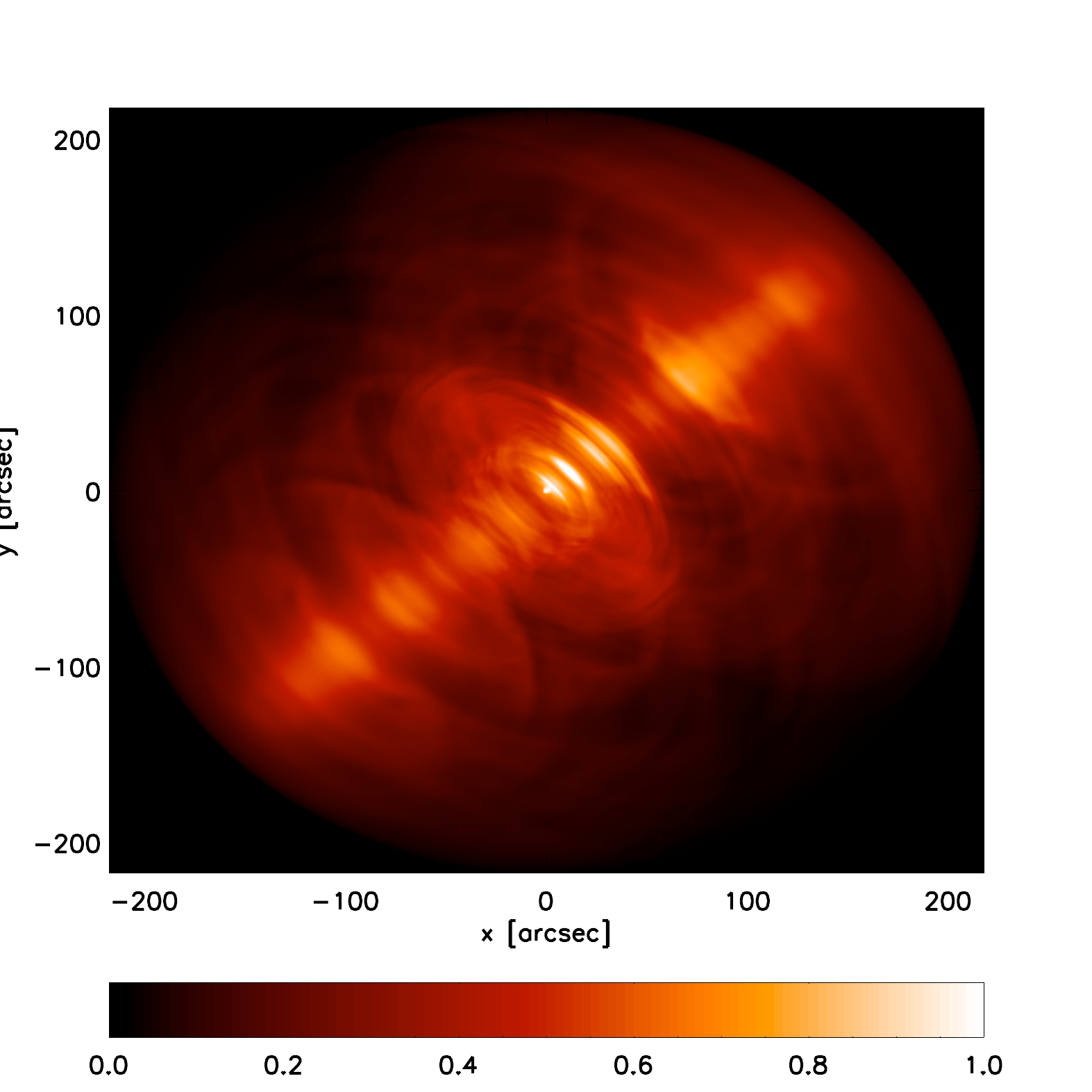}
\includegraphics[scale=.3]{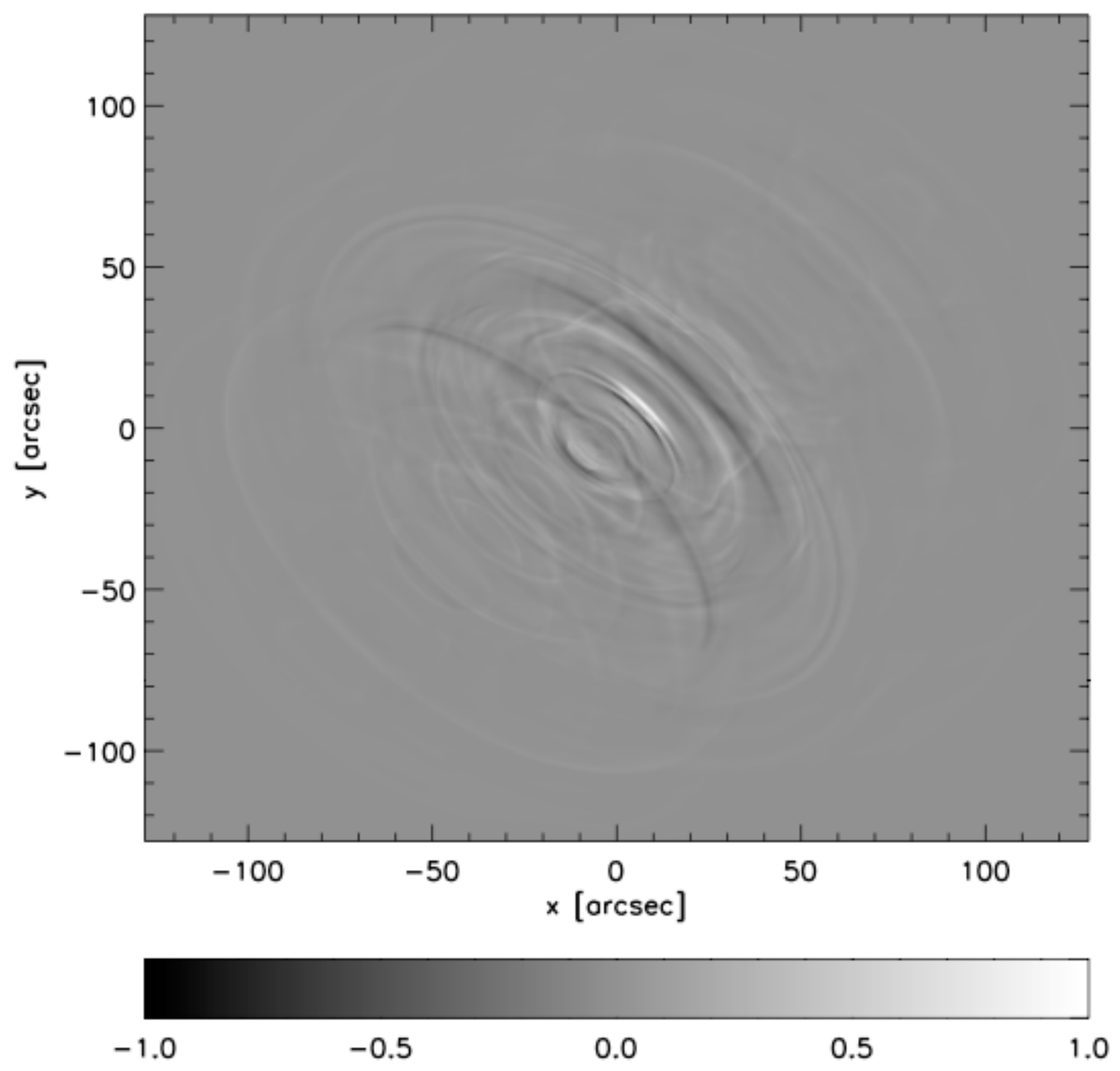}
\caption{Left panel: simulated emission map at a frequency of 1.4 GHz. The map is computed assuming a spatially uniform distribution of emitting particles. Right panel: changes in the inner nebula (radio "wisps") resulting from subtraction from the map on the left of a similar one computed for a 2 months later time.}
\label{fradio}
\end{figure}

When synchrotron emission is computed on top of the MHD simulations (see Fig.~\ref{fradio}), these two different hypotheses on the spatial distribution of the radio particles produce very similar results: the emission maps are very similar and comparison with observations does not allow to discriminate between the two cases. In addition, also time-variability in the inner region is found in both cases and with very similar features. This is an important result, because the existence of "radio wisps" could be naively taken as evidence supporting ongoing injection of radio particles at the TS, while Ref.~\refcite{olmi13} shows that this conclusion is not straightforward. Wisp variability arises in simulated maps from the properties of the MHD flow. The moving bright features are associated with locally enhanced magnetic field and Doppler boosting. The fact that the wisps are not coincident at different wavelengths \cite{bieten04,schweizer13} actually suggests a difference in the acceleration sites of the particles responsible for the emission. This could mean that radio emitting particles are accelerated somewhere else than the TS, or simply that particles of different energies (different sectors of the spectrum, which is a broken power-law) are accelerated at different latitudes along the shock front.

\section{Possible Acceleration mechanisms}
\label{sec:accel}
Let us finally briefly discuss what are the possibilities in terms of acceleration mechanisms to explain the broad band spectrum of the Crab Nebula. I already mentioned that the pulsar wind termination shock is in principle a hostile environment for particle acceleration and yet it accelerates particles with very high efficiency and up to very high energies. The resulting particle spectrum is a broken power-law, $N(E) \propto E^{-\gamma_p}$, with $\gamma_p \approx 1.5$ at low energies and $\gamma_p \approx 2.2$ at high energies.

\begin{figure}
\includegraphics[scale=.28]{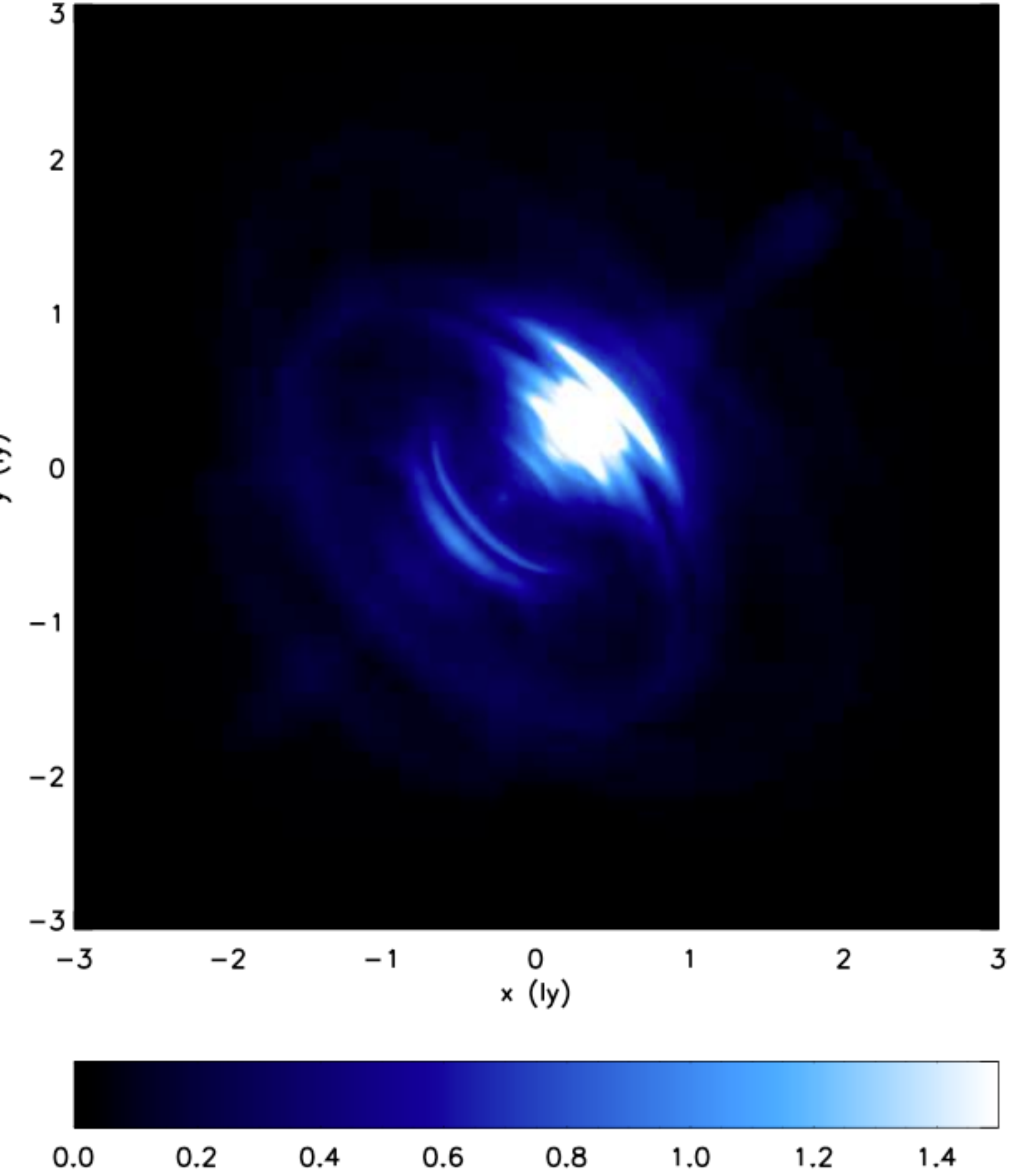}
\includegraphics[scale=.28]{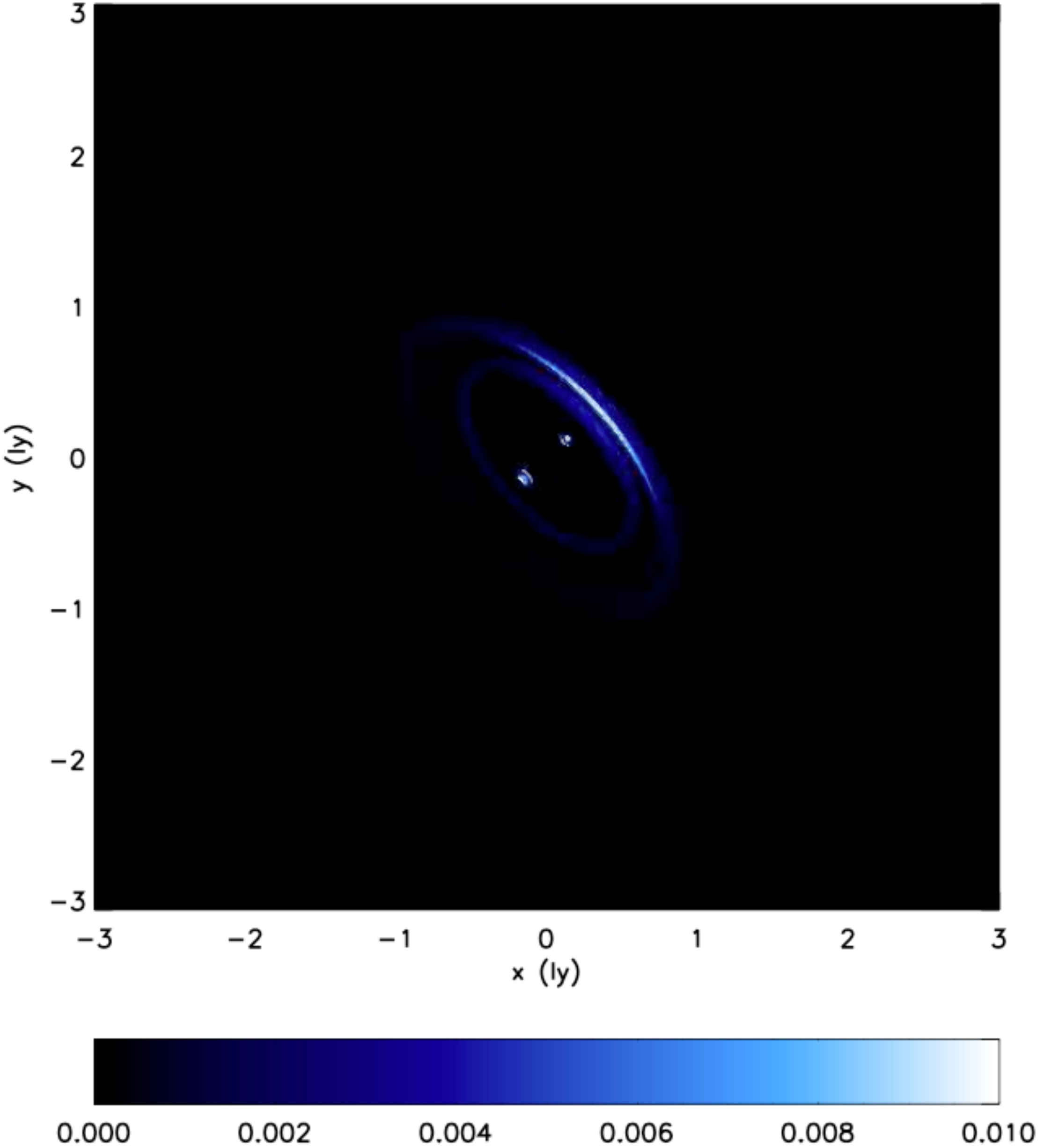}
\caption{Simulated X-ray emission from the Crab Nebula. Left panel: simulated X-ray emission. Right panel:
emission map including only particles that come from regions of the TS with $\sigma<10^{-3}$.}
\label{facc}
\end{figure}

The TS is a transverse relativistic shock in a flow with $\sigma>0.03$ on average. This implies that Fermi I type acceleration is not an option \cite{sironi09}: this mechanism is the most commonly invoked at shocks, and in addition would naturally give the slope of the X-ray emitting part of the spectrum. However, $\sigma$ is low enough as to allow Fermi I process to operate only on a very small fraction of the shock surface (remember that the TS is oblate and $\sigma$ depends on latitude). The fraction of flow energy through this sector of the TS is $\eta_{\rm low}< few \%$, and hence insufficient to explain the luminosity of the Crab Nebula (see Fig.~\ref{facc}). This estimate was obtained based on 2D simulations. 3D simulations seem to favor higher values of $\sigma$ \cite{porth13}, so it is likely that more realistic estimates will lead to an even lower $\eta_{\rm low}$.

Alternative proposals that have received some attention are: 1) driven magnetic reconnection at the TS; 2) resonant absorption of ion-cyclotron waves in a ion-doped plasma. The viability of both proposals depends on the wind composition and pulsar multiplicity. 
Driven reconnection has recently been investigated by Ref.~\refcite{sironi11} in a set-up appropriate to describe the Crab Nebula. The idea is that if the pulsar wind keeps its stripes with opposite direction of the magnetic field all the way to the TS, here compression causes the field to reconnect. A number of reconnection islands develop in the flow, where unscreened electric fields can effectively accelerate particles: the slope and the extension in energy of the resulting particle spectrum depend on the flow magnetization and on the ratio between the wavelength of the stripes and the particle Larmor radius. The latter ratio can be expressed in terms of $\kappa$ and the final result is that in order to reproduce the spectrum of radio particles, one would need $\sigma>30$ ad $\kappa >10^7$. Even ignoring the fact that such a high value of $\kappa$ is difficult to explain from the theory point of view \cite{arons12}, a wind with such a large number of pairs is actually likely to reconnect before reaching the TS \cite{kirkskja}, causing $\sigma$ to decrease below the required minimum. The requirements on $\kappa$ would be less severe if the process took place at high latitudes, since here the shock is closer to the pulsar and the particle density scales as $r^{-2}$. However at high latitudes one does not expect any stripes.

An alternative proposal, that works for whatever $\sigma$, but requires that most of the energy of the pulsar wind be carried by ions, is that of resonant absorption by the pairs of the cyclotron radiation emitted by such ions. The idea is that at the crossing of the TS, the sudden enhancement of the magnetic field sets the previously drifting plasma into gyration. The leptons quickly thermalize through emission and absorption of cyclotron waves, but ions with the same initial Lorentz factor (the wind is cold, so that all particles were moving with the same bulk Lorentz factor) react on time-scales that are longer by a factor $m_i/m_e$. If the wind is sufficiently cold ($\delta u/u < m_e/m_i$, with $u$ the four velocity) before the TS, the ions emit waves with large power not only at the fundamental frequency of their gyration, but up to a frequency $m_i/m_e$ times higher, which can then be resonantly absorbed by the pairs \cite{amatoarons}. The resulting acceleration efficiency $\epsilon_{\rm acc}$, spectral slope $\gamma_p$ and maximum energy $E_{\rm max}$, all depend on the fraction of energy carried by the ions $U_i/U_{\rm tot}$. PIC simulations show a wide variety of values:  $\epsilon_{\rm acc}=few (30)\%$, $\gamma_p >3 (<2)$, $E_{\rm max}/(m_i \Gamma c^2)=0.2(0.8)$ for $U_i/U_{\rm tot}=0.6(0.8)$. Once again the pulsar multiplicity and the related question of the origin of radio particles play a crucial role: if radio particles are part of the flow, and hence $\kappa \gg 10^4$, even if ions were extracted from the pulsar at the maximum possible rate, $\dot N_{GJ}$, they could not be energetically dominant and this mechanism is not an option.  

A cautionary remark is in order: a possibility is that even if $\kappa$ is overall very high, the composition of the wind might be latitude dependent and there might be sectors in which ions locally dominate the energy budget. However, once again the fact that the X-ray power radiated by the Crab Nebula is such a large fraction of $\dot E$ is very challenging: in terms of energy the acceleration process cannot be limited to a small fraction of the flow.

\section{Summary and conclusions}

Current 2D MHD models are very successful at explaining the high energy morphology of PWNe, but they are not free of problems. The main issue is that they require low $\sigma$ and lead to a nebular magnetic field which is too low to account for the combined synchrotron and ICS emission spectrum. A possibility to overcome these difficulties comes from the first 3D studies on the subject, which show that kink instabilities might play an important role in tangling the field, reducing the hoop stress and allowing to reproduce the morphology with a larger magnetic field strength. Current simulations however have too short a duration and probably also excessive dissipation. Further investigation is needed.

A conclusion that seems unavoidable, however, is that X-ray particles must be accelerated in high $\sigma$ regions and this poses a serious puzzle in terms of identifying their acceleration mechanism. The only process proven to work at a highly relativistic magnetized shock in a plasma that is mostly made of pairs is resonant absorption of the waves emitted by a heavier component, that is a minority by number but carries most of the energy. This is a viable mechanism only if/where the ratio between the number of pairs and that of ions does not exceed the mass ratio, a condition that translates into $\kappa < 10^4$.  
This requires in turn that the radio particles cannot be part of the current pulsar outflow. Indeed our investigation of the radio emission morphology and time variability does not exclude that radio emitting particles might come from elsewhere than the termination shock. In particular, while a scenario in which they do not come from the pulsar (but e.g. from the SN ejecta) is not really convincing (especially due to the similarities between PWNe and Bow Shock PWNe), the radio particles could be relics of earlier times re-energized by distributed turbulence in the nebula. If this were the case, cyclotron absorption could then provide acceleration of the higher energy part of the spectrum. 

Direct evidence that the acceleration sites of radio and X-ray emitting particles cannot be fully coincident comes from the observation that the wisps are not coincident at the different frequencies \cite{bieten04,schweizer13}. Further insight is likely to come from detailed modeling of this phenomenology.

\section*{Acknowledgments}
Thanks are due to J. Arons, R. Bandiera, N. Bucciantini, L. Del Zanna, B. Olmi and D. Volpi, my main collaborators on this subject during the years, and to F. Aharonian for financial support allowing me to participate to this conference.

\end{document}